\documentclass[conference]{IEEEtran}
\IEEEoverridecommandlockouts
\usepackage{listings}
\usepackage{xcolor}
\usepackage{booktabs}
\usepackage{graphicx}
\usepackage{multirow}
\lstdefinestyle{prompt}{
  basicstyle=\ttfamily\footnotesize,
  breaklines=true,
  frame=single,
  columns=fullflexible,
  aboveskip=6pt,
  belowskip=6pt
}
\lstdefinestyle{trace}{
  basicstyle=\ttfamily\scriptsize,
  breaklines=true,
  frame=single,
  columns=fullflexible,
  aboveskip=4pt,
  belowskip=4pt
}
\lstdefinestyle{verilog}{
  basicstyle=\ttfamily\footnotesize,
  breaklines=true,
  frame=single,
  columns=fullflexible,
  aboveskip=4pt,
  belowskip=4pt
}

\usepackage{cite}
\usepackage{amsmath,amssymb,amsfonts}
\usepackage{algorithmic}
\usepackage{graphicx}
\usepackage{textcomp}
\usepackage{xcolor}
\def\BibTeX{{\rm B\kern-.05em{\sc i\kern-.025em b}\kern-.08em
    T\kern-.1667em\lower.7ex\hbox{E}\kern-.125emX}}
\begin{document}

\title{Exploring the Agentic Frontier of \\ Verilog Code Generation}
\author{\IEEEauthorblockN{Patrick Yubeaton}
\IEEEauthorblockA{New York University\\
Tandon School of Engineering\\
Brooklyn, NY}
\and
\IEEEauthorblockN{Siddharth Garg}
\IEEEauthorblockA{New York University\\
Tandon School of Engineering\\
Brooklyn, NY}
\and
\IEEEauthorblockN{Chinmay Hegde}
\IEEEauthorblockA{New York University\\
Tandon School of Engineering\\
Brooklyn, NY}
}

\maketitle

\begin{abstract}
Large language models (LLMs) have made rapid advancements in code generation for popular languages such as Python and C++. Many of these recent gains can be attributed to the use of ``agents'' that wrap domain-relevant tools alongside LLMs. Hardware design languages such as Verilog have also seen improved code generation in recent years, but the impact of agentic frameworks on Verilog code generation tasks remains unclear. In this work, we present the first systematic evaluation of agentic LLMs for Verilog generation, using the recently introduced CVDP benchmark. We also introduce several open-source hardware design agent harnesses, providing a model-agnostic baseline for future work. Through controlled experiments across frontier models, we study how structured prompting and tool design affect performance, analyze agent failure modes and tool usage patterns, compare open-source and closed-source models, and provide qualitative examples of successful and failed agent runs. Our results show that naive agentic wrapping around frontier models can degrade performance (relative to standard forward passes with optimized prompts), but that structured harnesses meaningfully match and in some cases exceed non-agentic baselines. We find that the performance gap between open and closed source models is driven by both higher crash rates and weaker tool output interpretation. Our exploration illuminates the path towards designing special-purpose agents for verilog generation in the future.  
\end{abstract}

\begin{IEEEkeywords}
agents, benchmarking, CVDP, RTL generation
\end{IEEEkeywords}

\section{Introduction}

The rapid progress of LLMs in software code generation has been driven in large part by agentic frameworks: systems in which an LLM is paired with tools such as compilers, linters, file editors, and shell access to iteratively refine its outputs~\cite{yang2024swe, jimenez2024swebench}. These frameworks have proven highly effective for general-purpose software development tasks, witnessed by the explosion in real-world usage of agents such as Anthropic's Claude Code and OpenAI's Codex~\cite{plaat2025agentic}. However, the specialization of agents to hardware description languages (HDLs) such as Verilog remains largely unexplored~\cite{yang2025large}.

Verilog generation from specifications is a qualitatively harder problem than generic software code generation. Hardware designs must satisfy strict semantic correctness requirements across all possible input states, and errors that look plausible in isolation can cause timing violations or race conditions that only surface in late-stage verification. Existing LLM-based Verilog generation work has largely operated in non-agentic, single-shot settings~\cite{lu2024rtllm, liu2023verilogeval, yubeaton2025verithoughts}, leaving open the question of whether agentic tool use can provide the same gains for hardware design that it has for software engineering.

The recently introduced Comprehensive Verilog Design Problems (CVDP) benchmark~\cite{pinckney2025comprehensive} was specifically designed to address this gap. CVDP covers a wide range of Verilog generation and debugging tasks and is structured around two subsets: a non-agentic subset intended for single-pass evaluation, and an agentic subset intended for systems that can interact with a shell environment and iteratively refine solutions using compilation and simulation feedback. Despite being designed for agentic evaluation, as of the time of this manuscript's submission, CVDP has not yet been evaluated in its intended agentic setting.

In this paper, we make two primary contributions. First, we present the first open-source, hardware design-focused, model-agnostic agentic framework that equips any frontier LLM with Verilog-relevant tools including a compiler (iverilog)~\cite{williams2002icarus}, simulator (vvp), linter (Verilator), and synthesis tool (Yosys)~\cite{wolf2013yosys}. Second, we provide the first systematic evaluation of agentic LLMs on CVDP, spanning frontier commercial models and a range of agent configurations. Our experiments are organized around four research questions:

\begin{itemize}
    \item \textbf{RQ1:} How do frontier LLMs perform on CVDP in a non-agentic, single-pass setting?
    \item \textbf{RQ2:} Does wrapping a frontier LLM in an agentic framework with tool access improve or degrade performance on CVDP?
    \item \textbf{RQ3:} Can improvements to the agent's system prompt design and tool catalog further improve performance?
    \item \textbf{RQ4:} What are the primary failure modes of the agent, and which tool usage patterns are associated with correct versus incorrect outcomes?
\end{itemize}

Our results show that naive agentic wrapping often degrades performance relative to the non-agentic baseline, but that a more structured system prompt---one that enforces a fixed sequence of file discovery, planning, editing, and verification steps---recovers and in some cases exceeds non-agentic performance. Expanded tooling (Yosys, Verilator) provides only marginal additional benefit, suggesting that the bottleneck is model reasoning rather than tool availability. Failure mode analysis reveals that agent crashes are highly predictive of task failure, and that running simulation (\texttt{vvp}) is the tool usage pattern most strongly associated with correct outcomes.

\section{Background \& Related Work}

\subsection{Benchmarking LLM Verilog Generation}

Early benchmarks for LLM-based Verilog generation evaluated models in single-pass, non-agentic settings. RTLLM~\cite{lu2024rtllm} introduced an open-source benchmark for specification-to-RTL generation and demonstrated that general-purpose LLMs could produce plausible Verilog designs with appropriate prompting, though correctness remained a challenge for complex modules. VerilogEval~\cite{liu2023verilogeval} established a more rigorous evaluation framework using simulation-based testbenches drawn from the HDLBits platform, enabling pass@k evaluation of functional correctness across a range of design tasks. More recently, VeriThoughts~\cite{yubeaton2025verithoughts} introduced formal equivalence checking as a stricter correctness criterion, pairing a benchmark of specification-to-RTL tasks with Yosys-based logical equivalence checking against golden reference designs.

These benchmarks share a common limitation: they evaluate models in non-agentic, single-shot settings, without multi-turn interaction or access to external tools. CVDP~\cite{pinckney2025comprehensive} addresses this directly by providing a large-scale benchmark explicitly designed for agentic evaluation. CVDP covers a diverse set of Verilog generation, debugging, and verification tasks, and partitions problems into a non-agentic subset (solvable from context alone) and an agentic subset (requiring shell interaction, file editing, and tool use). Frontier models currently achieve relatively low scores on CVDP, making it a useful stress test for both models and agent designs. CVDP has been explored across a range of recent efforts~\cite{wu2026llmfsmscalinglargelanguage} - \nocite{kochar2026grpostatemutationsimproving, khan2026bench4hlsendtoendevaluationllms, lu2026newbenchmarkappropriateevaluation, zhang2026llm4covexecutionawareagenticlearning, huang2025assessinglargelanguagemodels, deng2026acertlagenticcontextevolution, ghorab2025notsotinylargelivingbenchmark, chen2026siliconmindv1multiagentdistillationdebugreasoning, tsai2025multimodalchipphysicaldesign, kumar2026saarthiagidomainspecificgeneral, chang2026specloopagenticrtltospecificationframework}\cite{lyu2026localvexploitinginformationlocality}. However, to our knowledge, no prior work has evaluated CVDP in its intended agentic setting.

\subsection{LLMs and Terminal Agents}

The use of LLMs as autonomous agents operating within terminal or shell environments has grown rapidly in software engineering. SWE-agent~\cite{yang2024swe} demonstrated that a purpose-built agent-computer interface---providing the LLM with structured commands for file navigation, editing, and execution---substantially improved performance on software engineering tasks over naive tool access. OpenHands ~\cite{wang2025openhandsopenplatformai} extended this paradigm to a more general multi-agent architecture, enabling LLMs to delegate subtasks and coordinate across multiple tool-using sub-agents.

In commercial deployments, Claude Code and OpenAI Codex represent widely used agentic coding systems~\cite{sobo2025evaluating}, providing LLMs with shell access, code execution, and file editing capabilities within iterative multi-turn loops. On the open-source side, Terminus~\cite{openthoughts-agent} offers a terminal-native coding agent that similarly equips LLMs with file system and execution tools. A consistent finding across these systems is that agent design---particularly the structure of the system prompt and the selection of available tools---has an outsized impact on performance, often exceeding the impact of model scale alone. These findings from software engineering motivate our investigation into whether similar design principles transfer to the hardware design domain.

\section{Creating a Verilog Agent}

Due to the lack of open-source hardware design agents, we follow best practices from the terminal agent community when designing our Verilog agent. Our goal is to have a simple, model-agnostic agent that can serve as a baseline against future agents designed or trained for specific models or problem types. The agent must fulfill a few core requirements:

\begin{enumerate}
    \item It must be model agnostic.
    \item It must provide the model with Verilog-specific tools as well as Linux commands necessary to complete CVDP tasks.
    \item It must allow the model to iterate upon its answer until it is ready to submit.
\end{enumerate}

Our agent operates in the following loop when solving a task:

\begin{enumerate}
    \item Read the prompt from CVDP. Provide the system message and prompt to the LLM.
    \item Receive a response from the LLM and parse it for tool calls.
    \item If tool calls are present, execute each and return the results as input to the LLM.
    \item Continue until the LLM signals that it has formulated a final answer.
\end{enumerate}

We note that the two major design choices within this framework are the system message and the tool catalog. Although CVDP does not provide a reference agent implementation, it does provide a baseline system message, which we adopt and modify as our starting point (see Appendix~\ref{app:baseline_prompt}). The baseline prompt gives the agent access to standard Linux commands for file reading and writing, as well as iverilog for compilation and vvp for simulation. 

Key design properties of the baseline system message include: (1) a tool catalog covering file I/O (\texttt{ls}, \texttt{cat}, \texttt{echo}, \texttt{sed}, \texttt{awk}), Verilog compilation (\texttt{iverilog}), and simulation (\texttt{vvp}); and (2) a lightweight thought/action/observation loop structure to organize reasoning steps. In several experiments below, we examine each of these design choices. 

\section{Experiments}

We evaluate on CVDP in three settings: the non-agentic subset without tooling, the agentic subset without tooling, and the agentic subset with tooling. The non-agentic and agentic subsets are defined by the CVDP dataset developers based on whether a problem is expected to require multi-turn interaction and tool use. Our experiments are organized around four research questions.

\subsection{RQ1: Non-Agentic Performance of Frontier LLMs}

In the non-agentic and non-tool agentic settings, models are given the full prompt, context (RTL files, specification files, etc.), and all other problem information in a single context window. The model is expected to reason through the problem and return its answer in one pass, without multi-turn interaction. This setting isolates raw model capability from agent scaffolding and serves as a baseline for RQ2.

Results across frontier models are shown in Table~\ref{tab:agentic_vs_nonagentic}. Performance on the non-agentic subset varies substantially across models, with Gemini-3.1 Pro Preview achieving the highest score at 58.61\%. Even the strongest models leave substantial room for improvement, consistent with CVDP's position as a difficult frontier benchmark.

\subsection{RQ2: Does Agentic Tool Use Help?}

We wrap the same frontier models in our baseline agent and evaluate on the agentic subset of CVDP, both with and without tool access. In the tool use setting, models receive only the initial prompt and must use Linux commands to discover additional files and context. Crucially, the model cannot submit its answer as plain text; it must modify the RTL file using shell commands and then signal completion. This imposes additional difficulty beyond the non-agentic setting.

As shown in Table~\ref{tab:agentic_vs_nonagentic}, performance in the tool-use agentic setting is consistently lower than the non-agentic baseline for most models. The no-tool agentic setting performs somewhat better than the tool-use setting in several cases, suggesting that the extra agent scaffolding itself---rather than tool access specifically---introduces failure modes. Possible explanations include models struggling to correctly format and sequence tool calls, or the added constraint of file-based submission causing errors.

\begin{table}[t]
\centering
\resizebox{\columnwidth}{!}{%
\begin{tabular}{lccc}
\toprule
\textbf{Model} & \textbf{Non-Agentic} & \textbf{Agentic (Tool)} & \textbf{Agentic (No Tool)} \\
\midrule
Gemini-3.1 Pro Preview & 58.61\% & 42.39\% & 47.39\% \\
GPT Codex-5.3          & 49.67\% & 45.65\% & 41.96\% \\
Claude Opus 4.6        & 50.66\% & 43.48\% & 36.74\% \\
KimiK-2.5              & 47.35\% & 21.74\% & 34.05\% \\
MiniMax                & 33.11\% & 23.55\% & 23.91\% \\
\bottomrule
\end{tabular}
}
\vspace{3pt}
\caption{Pass@1 performance across non-agentic and agentic evaluation settings.}
\label{tab:agentic_vs_nonagentic}
\end{table}

\subsection{RQ3: Can Structured Prompting and Expanded Tooling Improve the Agent?}

Given the performance gap observed in RQ2, we investigate whether improvements to the system message and tool catalog can recover or exceed non-agentic baselines. We introduce an updated agent with a more structured five-step system prompt (see Appendix~\ref{app:updated_prompt}) that enforces the following sequence for every problem:

\begin{enumerate}
    \item \textbf{Discover and read all files.} The agent must run \texttt{ls -R} and \texttt{cat} every file before any reasoning begins.
    \item \textbf{Plan changes explicitly.} The agent writes out a plan and justifies all intended edits before touching any file.
    \item \textbf{Apply changes.} The agent implements only the planned changes using shell commands.
    \item \textbf{Verify using all applicable tools.} The agent must achieve a successful \texttt{iverilog} compilation before proceeding. It also runs Verilator (semantic linting), Yosys lint and synthesis (structural checking), and formal verification if applicable. Critically, the agent is instructed to ignore warnings in pre-existing files it did not modify.
    \item \textbf{Signal completion.} The agent calls \texttt{task\_complete} only after all applicable checks pass.
\end{enumerate}

We also expand the tool catalog to include Yosys and Verilator alongside the baseline tools, to test whether broader verification coverage improves performance. Results are shown in Table~\ref{tab:updated_agent}.

\begin{table}[t]
\centering
\resizebox{\columnwidth}{!}{%
\begin{tabular}{lcccc}
\toprule
\textbf{Model} & \textbf{No Tooling} & \textbf{Baseline Agent} & \textbf{Updated Sys Msg} & \textbf{New Tooling} \\
\midrule
GLM-4.7                & 27.17\% & 28.26\% & 27.17\% & 28.26\% \\
Kimi K2.5              & 34.05\% & 21.74\% & 25.00\% & 23.91\% \\
Gemini-3.1 Pro Preview & 47.39\% & 42.39\% & 47.61\% & 47.39\% \\
Gemini-2.5 Flash       & 17.39\% & 16.30\% & 20.65\% & 22.83\% \\
\bottomrule
\end{tabular}
}
\vspace{3pt}
\caption{Pass@1 results comparing different agent configurations on the agentic CVDP subset.}
\label{tab:updated_agent}
\end{table}

The updated system prompt yields consistent improvements over the baseline agent across all tested models, with Gemini-3.1 Pro Preview recovering to match its no-tooling baseline (47.61\% vs. 47.39\%). In contrast, adding expanded tooling (Yosys, Verilator) provides only minimal further benefit. This suggests that the performance bottleneck lies in the model's reasoning and planning behavior rather than in the breadth of verification tools available. The structured step-by-step prompt appears to reduce failure modes related to premature submission and insufficient file exploration.

\subsection{RQ4: Agent Failure Modes and Tool Usage Patterns}

Having established that structured prompting improves aggregate pass rates, we now examine \emph{why} agents fail and \emph{which tool usage patterns} differentiate correct from incorrect runs. We focus this analysis on the four models for which we have data across all three agent configurations: Gemini 3.1 Pro, Gemini 2.5 Flash, Kimi K2.5, and GLM-4.7.

\subsubsection{Agent Completion and Crash Rates}

Table~\ref{tab:completion_crash} summarizes agent completion and crash rates across all model-configuration pairs. Two patterns stand out. First, agent completion rates vary dramatically across models: Gemini 3.1 Pro achieves 90.9\% completion under the baseline agent, whereas Gemini 2.5 Flash, Kimi K2.5, and GLM-4.7 complete only 64.1\%, 60.9\%, and 72.8\% of runs respectively. Second, the updated system prompt (Mod 1) dramatically reduces crash rates for all models. Gemini 3.1 Pro's crash rate drops from 9.1\% to 1.7\%; GLM-4.7's drops from 27.2\% to 12.0\%; and Kimi K2.5's drops from 39.1\% to 25.0\%. The new tooling configuration (Mod 2) shows a mixed picture: it further reduces crashes for GLM-4.7 (to 8.7\%) and Gemini 3.1 Pro (to 4.1\%), but does not consistently help or hurt Kimi K2.5 or Gemini 2.5 Flash.

\begin{table}[t]
\centering
\resizebox{\columnwidth}{!}{%
\begin{tabular}{llcc}
\toprule
\textbf{Model} & \textbf{Config} & \textbf{Completed} & \textbf{Crashed} \\
\midrule
\multirow{3}{*}{Gemini 3.1 Pro}  & Baseline & 90.9\% & 9.1\% \\
                                   & Mod 1    & 98.3\% & 1.7\% \\
                                   & Mod 2    & 95.9\% & 4.1\% \\
\midrule
\multirow{3}{*}{Gemini 2.5 Flash} & Baseline & 64.1\% & 35.9\% \\
                                   & Mod 1    & 68.5\% & 31.5\% \\
                                   & Mod 2    & 60.9\% & 39.1\% \\
\midrule
\multirow{3}{*}{Kimi K2.5}        & Baseline & 60.9\% & 39.1\% \\
                                   & Mod 1    & 75.0\% & 25.0\% \\
                                   & Mod 2    & 64.1\% & 35.9\% \\
\midrule
\multirow{3}{*}{GLM-4.7}          & Baseline & 72.8\% & 27.2\% \\
                                   & Mod 1    & 88.0\% & 12.0\% \\
                                   & Mod 2    & 91.3\% & 8.7\% \\
\bottomrule
\end{tabular}
}
\vspace{3pt}
\caption{Agent completion and crash rate across models and configurations.}
\label{tab:completion_crash}
\end{table}

\subsubsection{Failure Mode Taxonomy}

We see four distinct failure modes. The dominant mode across all models is \texttt{unknown} (the agent loop completes but the solution is wrong), which accounts for 50--96\% of failures depending on the model and configuration. The \texttt{no\_log} / \texttt{agent\_crash} modes (which co-occur and represent hard agent failures) are the second most common, and are substantially reduced by the updated system prompts. \texttt{harness\_fail} (evaluation infrastructure failure) is rare ($\leq$2\% across all configurations) and can be treated as noise.

The persistence of \texttt{unknown} failures even under the best agent configuration indicates that the primary bottleneck is correctness of the generated Verilog itself, not the agent's ability to navigate the task structure. This is consistent with the finding in RQ3 that expanded tooling provides minimal benefit: the tools can confirm compilation success but cannot guarantee functional correctness.

\subsubsection{Tool Usage and Correctness}

Table~\ref{tab:tool_delta} reports the tools with the largest positive and negative deltas in usage rate between correct and incorrect runs for Gemini 3.1 Pro, the model with the most data, across all three configurations. Several patterns emerge consistently:

\textbf{Simulation (\texttt{vvp}) is the strongest positive signal.} In the baseline agent, \texttt{vvp\_direct} is used in 27.2\% of correct runs versus only 7.8\% of incorrect runs (+19.3 pp delta). This pattern holds under Mod 1 (\texttt{vvp\_simulate}: 64.2\% correct vs. 40.9\% incorrect, +23.3 pp) and Mod 2 (\texttt{vvp\_simulate}: 61.0\% vs. 34.7\%, +26.3 pp). Running simulation to verify functional behavior is the single most discriminating signal between passing and failing runs---consistent with the expectation that the model is more likely to catch errors when it actually exercises its design.

\textbf{Compilation (\texttt{iverilog}) is near-universal but shows modest positive signal.} Under the baseline, \texttt{iverilog\_direct} is used in 80.7\% of correct versus 72.4\% of incorrect runs (+8.3 pp). This positive delta is expected: runs that compile are more likely to be structurally valid. However, compilation success does not guarantee functional correctness, so the delta is modest.

\textbf{Excessive \texttt{sed} usage is negatively associated with correctness.} In both the baseline and Mod 2 configurations, \texttt{sed} is used more frequently in incorrect runs: baseline $-11.6$ pp delta, Mod 2 $-13.5$ pp. This is consistent with the agent making multiple in-place file edits, possibly indicating repeated failed repair attempts.

\textbf{Excessive \texttt{find} and \texttt{fs\_ops} usage is negatively associated.} Both the \texttt{find} ($-11.1$ pp baseline, $-4.3$ pp Mod 1, $-3.7$ pp Mod 2) and \texttt{fs\_ops} ($-6.9$ pp baseline, $-13.3$ pp Mod 1, $-8.6$ pp Mod 2) tools show consistent negative deltas. Runs that rely heavily on filesystem search operations may indicate the agent is lost---spending turns searching for files rather than making progress on the solution.

\begin{table}[t]
\centering
\resizebox{\columnwidth}{!}{%
\begin{tabular}{llccc}
\toprule
\textbf{Tool} & \textbf{Signal} & \textbf{Baseline $\Delta$} & \textbf{Mod 1 $\Delta$} & \textbf{Mod 2 $\Delta$} \\
\midrule
\texttt{vvp} (simulate)    & Positive & +19.3 pp & +23.3 pp & +26.3 pp \\
\texttt{iverilog} (compile) & Positive & +8.3 pp  & +0.6 pp  & +0.1 pp  \\
\texttt{ls}                & Positive & +6.3 pp  & --       & --       \\
\midrule
\texttt{sed}               & Negative & $-$11.6 pp & $-$6.7 pp  & $-$13.5 pp \\
\texttt{find}              & Negative & $-$11.1 pp & $-$4.3 pp  & $-$3.7 pp  \\
\texttt{fs\_ops}           & Negative & $-$6.9 pp  & $-$13.3 pp & $-$8.6 pp  \\
\bottomrule
\end{tabular}
}
\vspace{3pt}
\caption{Tool usage rate delta (correct\% $-$ wrong\%) for Gemini 3.1 Pro across agent configurations. Positive values indicate more frequent use in correct runs; negative values indicate more frequent use in incorrect runs.}
\label{tab:tool_delta}
\end{table}

\section{Open-Source vs. Closed-Source Models}
\label{sec:open_closed}

The models in our agentic evaluation span two categories: closed-source frontier models (Gemini 3.1 Pro, Gemini 2.5 Flash)~\cite{team2023gemini} and open-source models (Kimi K2.5~\cite{team2026kimi}, GLM-4.7~\cite{glm2024chatglm}). We compare these groups across agent configurations to understand whether the performance gap observed in aggregate is driven by task difficulty scaling, tool usage quality, or both.

\subsection{Performance Gap Across Difficulty Levels}

Table~\ref{tab:open_closed_diff} shows pass rates and crash rates broken down by task difficulty for both model groups under the baseline agent. On easy tasks, all models perform comparably: pass rates range from 62.5\% to 71.4\%, suggesting that open-source models are capable of handling well-specified, low-complexity Verilog generation at near-parity with closed-source models.

\begin{table}[t]
\centering
\resizebox{\columnwidth}{!}{%
\begin{tabular}{llcccc}
\toprule
\textbf{Model} & \textbf{Type} & \multicolumn{2}{c}{\textbf{Medium}} & \multicolumn{2}{c}{\textbf{Hard}} \\
 & & Pass\% & Crash\% & Pass\% & Crash\% \\
\midrule
Gemini 3.1 Pro  & Closed & 48.7\% &  8.6\% & 51.2\% &  9.9\% \\
Gemini 2.5 Flash & Closed & 15.7\% & 41.2\% &  8.0\% & 36.0\% \\
Kimi K2.5       & Open   & 32.1\% & 45.3\% & 36.0\% & 40.0\% \\
GLM-4.7         & Open   & 43.4\% & 24.5\% & 28.0\% & 48.0\% \\
\bottomrule
\end{tabular}
}
\vspace{3pt}
\caption{Pass rate and crash rate by difficulty (medium and hard) for open and closed source models under the baseline agent. Easy tasks are omitted as all models achieve 62--71\% pass rates with comparable crash rates.}
\label{tab:open_closed_diff}
\end{table}

The gap opens sharply on medium and hard tasks. Gemini 3.1 Pro maintains 48.7\% and 51.2\% pass rates on medium and hard tasks with crash rates below 10\%. In contrast, Kimi K2.5 crashes on 45.3\% of medium runs and 40.0\% of hard runs, and GLM-4.7 crashes on 48.0\% of hard runs. The structured system prompt (Mod 1) substantially reduces crash rates for both open-source models---GLM-4.7's hard crash rate drops from 48.0\% to 24.0\%, and Kimi K2.5's from 40.0\% to 32.0\%---but does not close the underlying pass rate gap.

\subsection{Tool Usage Quality}

A key hypothesis is that open-source models may struggle to use tool outputs effectively, not just to avoid crashing. Table~\ref{tab:open_closed_tools} compares selected tool usage metrics across models under the baseline agent.

\begin{table}[t]
\centering
\resizebox{\columnwidth}{!}{%
\begin{tabular}{lcccc}
\toprule
\textbf{Tool} & \textbf{Gem 3.1 Pro} & \textbf{Gem 2.5 Flash} & \textbf{Kimi K2.5} & \textbf{GLM-4.7} \\
\midrule
\texttt{iverilog} usage\%  & 76.7\% & 34.8\% & 78.3\% & 100.0\% \\
\texttt{iverilog} avg/run  & 1.32   & 1.78   & 2.05   & 5.96 \\
\texttt{vvp} usage\%       & 18.0\% & 29.3\% & 66.3\% & 96.7\% \\
\texttt{vvp} avg/run       & 0.21   & 0.97   & 1.07   & 2.67 \\
\texttt{diff} usage\%      & 79.6\% & 17.4\% & 12.0\% & 21.7\% \\
\texttt{find} usage\%      & 13.5\% &  3.3\% & 48.9\% &  27.2\% \\
\bottomrule
\end{tabular}
}
\vspace{3pt}
\caption{Tool usage rates and average calls per run for open and closed source models under the baseline agent.}
\label{tab:open_closed_tools}
\end{table}

Two patterns are notable. First, GLM-4.7 invokes \texttt{iverilog} in 100\% of runs (averaging 5.96 calls per run) and \texttt{vvp} in 96.7\% of runs (2.67 calls per run)---far more than Gemini 3.1 Pro (1.32 and 0.21 respectively). Despite this high tool activity, GLM-4.7 passes only 43.5\% of problems, and its \texttt{finish\_to\_prob\_pass} rate of 55.2\% is only marginally better than chance. The repetitive compilation and simulation calls suggest the model is running tools in loops without effectively interpreting their feedback to improve the design.

Second, Kimi K2.5 uses \texttt{find} in 48.9\% of runs (avg. 0.54 calls), compared to 13.5\% for Gemini 3.1 Pro. As established in RQ4, \texttt{find} usage is a consistent negative signal associated with the agent being lost in the file system rather than making productive edits. This is a tool navigation failure rather than a Verilog reasoning failure.

These results suggest that the open vs. closed source performance gap on CVDP agentic tasks is driven by two compounding factors: (1) higher crash rates on complex tasks, particularly medium and hard difficulty problems, which closed-source models avoid through more robust instruction following; and (2) lower solution quality even in completed runs, potentially reflecting weaker tool output interpretation. Improving open-source model performance in agentic hardware settings likely requires both better crash robustness and better training on tool-use feedback loops.

\section{Qualitative Agent Examples}
\label{sec:qualitative}

To complement the aggregate statistics, we present two concrete examples that illustrate how the agent behaves on CVDP tasks: a successful run and a failed run due to overflow.

\subsection{Successful Run: Binary-to-Gray Converter}

Figure~\ref{fig:btg_prompt} shows the CVDP prompt for a parameterized binary-to-Gray code converter task. The prompt is concise, but additional detail is provided in a \texttt{docs/Spec.md} file that the agent must discover and read. This is a representative CVDP task structure: the prompt alone is insufficient to fully specify the design.

Figure~\ref{fig:btg_trace} shows an abbreviated version of the agent's tool call trace for this task under the updated system prompt (Mod 1). The agent follows the five-step structure: it first reads the spec and testbench, then writes the RTL, and then runs the full verification chain (\texttt{iverilog}, \texttt{verilator}, \texttt{yosys\_lint}, \texttt{yosys\_synth}, \texttt{get\_module\_ports}) before running simulation to confirm functional correctness. Critically, when Verilator reports timing-related warnings in the \emph{testbench} (a pre-existing file the agent did not modify), the agent correctly ignores them and reruns Verilator targeting only its own RTL file, obtaining a clean result.

\begin{figure}[t]
\centering
\begin{lstlisting}[style=trace]
[TOOL] shell_exec: 'ls -R'       -> discovers docs/, rtl/, verif/
[TOOL] shell_exec: 'cat ./docs/Spec.md'   -> reads full spec
[TOOL] shell_exec: 'cat ./verif/tb_binary_to_gray.sv' -> reads TB
[TOOL] shell_exec: cat << 'EOF' > ./rtl/binary_to_gray.sv
  module binary_to_gray #(parameter WIDTH = 6)(
      input  logic [WIDTH-1:0] binary_in,
      output logic [WIDTH-1:0] gray_out
  );
      assign gray_out = binary_in ^ (binary_in >> 1);
  endmodule
  EOF
[TOOL] iverilog_compile: ...     -> Compilation successful.
[TOOL] verilator_lint: (on TB)   -> timing warnings in TB (ignored)
[TOOL] verilator_lint: (RTL only)-> Lint clean.
[TOOL] yosys_lint: ...           -> Yosys lint clean.
[TOOL] yosys_synth: ...          -> Synthesis clean.
[TOOL] get_module_ports: ...     -> ports match spec.
[TOOL] shell_exec: 'iverilog ... && vvp a.out'
         -> Time=0 Binary=0000 Gray=0000
            Time=10 Binary=0001 Gray=0001 ...
[TOOL] task_complete called.
\end{lstlisting}
\caption{Abbreviated agent trace for the binary-to-Gray task (Gemini 3.1 Pro, Mod 1). The agent correctly ignores testbench warnings and verifies its own RTL independently.}
\label{fig:btg_trace}
\end{figure}

The resulting RTL, shown in Figure~\ref{fig:btg_rtl}, is a single-line combinational assignment implementing the standard Gray code formula ($\text{gray} = \text{binary} \oplus (\text{binary} >> 1)$), which is both concise and correct. This example illustrates the ideal agentic workflow: structured file discovery, concise correct RTL generation, and a multi-stage verification pass.

\begin{figure}[t]
\centering
\begin{lstlisting}[style=verilog]
module binary_to_gray #(
    parameter WIDTH = 6
)(
    input  logic [WIDTH-1:0] binary_in,
    output logic [WIDTH-1:0] gray_out
);
    assign gray_out = binary_in ^ (binary_in >> 1);
endmodule
\end{lstlisting}
\caption{RTL produced by the agent for the binary-to-Gray task. A single combinational assignment correctly implements the Gray code conversion.}
\label{fig:btg_rtl}
\end{figure}

\subsection{Failure Case: Context Overflow on a Complex Task}

Figure~\ref{fig:ca_fail} illustrates a hard failure mode: a context window overflow on the \texttt{cellular\_automata} pseudorandom generator task. The agent successfully generates an initial RTL implementation, iterates to fix an \texttt{iverilog} compatibility issue (replacing an \texttt{always\_comb} loop with a \texttt{generate} block), and runs simulation, which produces hundreds of lines of cycle-by-cycle output. That large output is fed back into the context, growing it until the next LLM call exceeds the model's 1M token limit, causing a hard crash with a \texttt{400 INVALID\_ARGUMENT} API error.

\begin{figure}[t]
\centering
\begin{lstlisting}[style=trace]
[TOOL] shell_exec: 'cat docs/specs.md'   -> reads CA spec
[TOOL] shell_exec: cat << EOF > pseudoRandGenerator_ca.sv
  ... (generates full CA module with Rule 30 / Rule 110) ...
  EOF && iverilog -g2012 ...
         -> ERROR: constant selects in always_* not supported
[TOOL] shell_exec: (rewrites with generate block instead)
         -> Compilation clean
[TOOL] shell_exec: 'iverilog ... && vvp a.out'
         -> Cycle 6074: Value 619b ... Cycle 6080: ...
            (hundreds of lines of simulation output ...)
[next API call]
google.genai.errors.ClientError: 400 INVALID_ARGUMENT.
  'The input token count exceeds the maximum number of
   tokens allowed 1048576.'
\end{lstlisting}
\caption{Abbreviated agent trace for the cellular automata task (Gemini 3.1 Pro, baseline). Large simulation output grows the context past the 1M token limit, causing a hard crash.}
\label{fig:ca_fail}
\end{figure}

This failure mode---verbose tool output flooding the context---is distinct from logical errors in the generated RTL. The agent produces functionally reasonable Verilog and correctly fixes a toolchain compatibility issue, yet crashes due to an infrastructure limitation rather than a reasoning failure. This motivated some of the tooling changes made in the Mod 1 agent and motivates future design to consider potential scaffolding issues.

\section{Discussion}

Our results demonstrate that agentic frameworks do not automatically improve Verilog generation performance: naively wrapping a frontier LLM in a tool-enabled agent can degrade performance relative to a single-pass baseline. However, structured system prompt design meaningfully closes this gap. This mirrors findings from the software engineering agent literature, where agent-computer interface design has been shown to be as impactful as model scale~\cite{yang2024swe}.

The failure mode analysis (RQ4) provides a clearer picture of where agents break down. Agent crashes are highly predictive of task failure, but the majority of failures occur in runs where the agent \emph{does} complete---the solution is simply wrong. This confirms that the bottleneck is not the agent scaffolding itself but the underlying model's Verilog reasoning ability. The strong positive association between simulation (\texttt{vvp}) usage and correct outcomes suggests that models which actually run their designs---rather than relying solely on static editing---are more likely to identify and fix errors before submission. The qualitative examples (Section~\ref{sec:qualitative}) illustrate both the ideal workflow and a concrete failure mode (context overflow) that is orthogonal to RTL correctness.

The open vs. closed source comparison (Section~\ref{sec:open_closed}) reveals that the performance gap widens substantially on medium and hard tasks, driven by two factors: higher crash rates for open-source models on complex problems, and lower solution quality even in completed runs. The structured system prompt reduces crashes but does not uniformly improve solution quality for open-source models, suggesting that prompt engineering alone is insufficient and that training-time improvements to tool-use capability are needed.

A key observation is that expanded tooling (Yosys, Verilator) provides only marginal benefit over a well-structured prompt with basic tools. This suggests that the primary bottleneck is internal Verilog reasoning ability rather than the breadth of available verification tools. Future work should explore training-time adaptation---fine-tuning or reinforcement learning against tool feedback---as a path toward agents that can more effectively leverage the full verification tool stack.

The CVDP benchmark deserves further attention as a resource. Its combination of diverse task types, agentic structure, and difficulty  makes it uniquely suited for evaluating the next generation of hardware design agents. We hope our open-source agent and evaluation serve as a reproducible baseline for future work in this space.

\section{Conclusion}

We introduced an open-source, model-agnostic Verilog agent and demonstrated that structured system prompt design is the key lever for improving agent performance, while expanded tooling provides only marginal gains. Failure mode analysis shows that agent crashes strongly predict task failure, that simulation usage is the tool pattern most associated with correct outcomes, and that the dominant failure mode even in well-functioning agents is incorrect Verilog generation---confirming that model reasoning, not agent scaffolding, is the primary bottleneck. These results establish a foundation for future work on training-time adaptation and more capable hardware design agents.
\newpage
\bibliographystyle{IEEEtran}
\bibliography{references}

\newpage
\appendix

\section{Baseline System Prompt}
\label{app:baseline_prompt}

\begin{figure}[h]
\centering
\begin{lstlisting}[style=prompt]
You are a language model that has the following file operations available at your disposal:

  - List files in a directory by running one of the following commands:
      - `ls`
      - `tree`

  - Read files by using:
      - `cat <filename>`

  - Write files by using:
      - `echo <content> > <filename>`

  - Compile Verilog by using `iverilog`, such as:
      - `iverilog -o <output_filename>.out -g2012 <verilog_code_file> <verilog_testbench_file>`

  - Run simulation by using:
      - `vvp <output_filename>.out`

  - Find the current working directory by using:
      - `pwd`

  - Update the contents of a text file from old content to new content:
      - `sed -i "problematic_line_number s/problematic_statement/non_problematic_statement/" Buggy_RTL_code.sv`

  - Access a specific line of a file:
      - `awk 'NR==line_number' file_name.sv`

You will be given a prompt, and your task is to understand it and solve the issue by using the above commands as needed. In the final step, you should create a Linux patch to highlight the necessary file updates to achieve the targeted goal.

You will solve the problem step by step using the following structure:

  - thought (the reasoning process for the step you are going to take)
  - action (the command you will run)
  - observation (the output from the action)

The last step will contain the final output summary and the patch itself in the following format:

  - thought (a summary of what you did and an introduction to the patch file)
  - patch (a Linux-based patch that needs to be applied to reach the relevant solution)
\end{lstlisting}
\caption{Baseline system prompt, adapted from the CVDP benchmark.}
\label{fig:system_prompt}
\end{figure}

\section{Updated System Prompt}
\label{app:updated_prompt}

\begin{figure}[h]
\centering
\begin{lstlisting}[style=prompt]
You are a Verilog hardware design assistant. Your task is to analyze, debug, or generate
Verilog/SystemVerilog code based on a given prompt.

You MUST follow this exact sequence of steps, do not skip or reorder them:

STEP 1: Discover and read all files
Run `ls -R` to list every file in the working directory. Then use `cat <filename>` to read
EVERY file you find (source files, testbenches, specs, READMEs, etc.). Do not proceed until
you have read all files in full.

STEP 2: Plan your changes
Think carefully about what edits or new files are required to satisfy the prompt.
Write out your plan explicitly before touching any file.
IMPORTANT: stay strictly within what the prompt and the files specify.
Do NOT infer extra requirements, add unrequested features, or change anything not directly
called for by the prompt or the existing specifications.

STEP 3: Apply changes
Implement your plan from Step 2 by modifying or creating files with Linux commands
(`sed`, `echo`, `awk`, `tee`, `cp`, `mv`, etc.).
Make only the changes you planned in Step 2.

STEP 4: Verify your implementation
Run all applicable verification tools in this order. Each tool targets different bug classes;
use ALL of them, not just the first one that passes:

  4a. `iverilog_compile`: confirms the RTL is syntactically valid.
  4b. `verilator_lint`: catches semantic issues iverilog misses.
  4c. `yosys_lint`: catches structural issues such as undriven outputs and port mismatches.
  4d. `yosys_synth`: catches synthesis-time issues including unintended latches.
  4e. `get_module_ports`: confirms port names, directions, and widths match the spec.
  4f. `formal_verify`: if assertions are present, run bounded model checking.

CRITICAL: Ignore warnings in pre-existing files you did NOT modify.
If any tool reports an error in your changed files, return to Step 2 and revise.

STEP 5: Signal completion
Once all applicable tools pass, call `task_complete` with a brief summary.
Do not call `task_complete` before a successful `iverilog_compile`.

At each step, structure your reasoning as:
  - thought     : what you are about to do and why
  - action      : the tool call / command
  - observation : the result
\end{lstlisting}
\caption{Updated structured system prompt with five-step verification loop.}
\label{fig:system_prompt_updated}
\end{figure}

\begin{figure}[t]
\centering
\begin{lstlisting}[style=prompt]
Design a `binary_to_gray` module in SystemVerilog. Refer to
the specification in `docs/specs.md`, which details a
parameterized `WIDTH` for an N-bit binary-to-Gray code
converter. The module should take an N-bit binary input and
generate an N-bit Gray code output using a purely
combinational approach. The design must follow the standard
Gray code conversion rule where:
  - The MSB remains unchanged.
  - Each subsequent bit is computed as the XOR of the
    current and previous binary bits.
Requirements:
  - Implement using bitwise XOR.
  - Ensure a fully combinational design (no clock or reset).
  - The module must be parameterized to support different
    bit widths.
\end{lstlisting}
\caption{CVDP prompt for the binary-to-Gray converter task.}
\label{fig:btg_prompt}
\end{figure}

\end{document}